# Generalized laws of reflection and refraction from transformation optics

Yadong Xu [1], Kan Yao [1], and Huanyang Chen [1, a)]

[1] School of Physical Science and Technology, Soochow University, Suzhou 215006, China

**Abstract:** Based on transformation optics, we introduce another set of generalized laws of reflection and refraction (differs from that of [Science **334**, 333 (2011)]), through which a transformation media slab is derived as a meta-surface, producing anomalous reflection and refraction for all polarizations of incident light.

The concept of phase discontinuities has inspired some thoughts on controlling light [1,2]. The key technique is to print V-shaped antennas on a dielectric surface. The antennas will excite cross-polarized scattered light with abrupt phase shifts from incident light. If the phase shifts are positionally dependant, the cross-polarized scattered light will obey a set of generalized laws of reflection and refraction, with a term of a gradient in a phase discontinuity added, whereas the scattered light with the same polarization obeys the conventional laws of reflection and refraction. In other words, using the V-shaped antennas, the excited cross-polarized scattered light is controllable, while the scattered light with the original polarization is still uncontrollable. Can we have a parallel phenomenon for all polarizations of light? The positionally dependant phase shifts suggest the likelihood for transformation optics to make this happen, as devices from transformation optics also possess positionally dependant material parameters [3,4]. Furthermore, the devices from transformation optics can control all polarizations of light when the required material parameters are fulfilled. Here in this letter, we will revisit the aforementioned phenomenon from the perspective of transformation optics. Another set of generalized laws of reflection and refraction will be obtained, with anomalous reflection and refraction achieved. Moreover, two more critical angles are derived for reflection and refraction respectively, which come from the transformation geometry itself.

The coordinate transformation is pretty simple, as shown in schematic plot in Fig. 1(a)

---

[a)] Author to whom correspondence should be addressed. Electronic mail: chy@suda.edu.cn

and Fig. 1(b),

$$x' = x, \quad x \geq 0,$$

$$y' = \frac{(y+h)h}{(y_0+h)} - h, \quad -h \leq y \leq y_0, \tag{1}$$

$$y_0 = kx = x\tan\theta_0.$$

By keeping the $x$ coordinate unchanged, a trapezoidal region (virtual space, Fig. 1(a)) is compressed into a rectangular one (physical space, Fig. 1(b)), where the boundaries $y=-h$ $(x\geq 0)$ and $x=0$ $(-h\leq y\leq 0)$ are fixed, while the boundary $y=kx$ $(x\geq 0)$ is compressed to $y'=0$ $(x'\geq 0)$ along each $x$ coordinate. Firstly, suppose that the whole virtual space is filled with a dielectric $n_1$, light will propagate in straight rays in it. From transformation optics, the material parameters in the slab $-h\leq y'\leq 0$ $(x'\geq 0)$ are

$$\overline{\overline{\varepsilon}} = \overline{\overline{\mu}} = n_1 \begin{bmatrix} 1/b & a/b & 0 \\ a/b & (a^2+b^2)/b & 0 \\ 0 & 0 & 1/b \end{bmatrix}, \tag{2}$$

where $a = \frac{\partial y'}{\partial x} = -(y+h)hk/(kx+h)^2$, $b = h/(kx+h)$. In the physical space, light will be bended or squeezed in the transformation media slab. After leaving the slab, its direction has a rotation at an angle of $\theta_0$. If the region beyond the boundary $y=kx$ $(x\geq 0)$ is filled with another dielectric $n_2$, there will be a reflection and a refraction at the interface. Suppose that the width of the slab $h$ is small enough for the interface to be regarded as a meta-surface. In the physical space, the laws of reflection and refraction should be modified as follows:

$$n_1 \sin(\theta_i + \theta_0) = n_2 \sin\theta_t, \tag{3}$$

$$\sin\theta_i = \sin(\theta_r - 2\theta_0), \tag{4}$$

where $\theta_i$ is the incident angle, $\theta_t$ is the refraction angle, $\theta_r$ is the reflection angle. This is another set of generalized laws of reflection and refraction. For the reflected light, as it passes the transformation media slab twice, the rays should have a direction rotation with an angle of

$2\theta_0$. The geometry of transformation automatically causes a critical angle of refraction ($\theta_{tc0}$) and a critical angle of reflection ($\theta_{rc0}$). For example, if the incident light is parallel to the boundary $y = kx$ ($x \geq 0$) (i.e., the incident angle is $90° - \theta_0$), physically it cannot reach the boundary $y' = 0$ ($x' \geq 0$). Such an angle is the intrinsic critical angle of refraction ($\theta_{tc0}$). Any light with an incident angle larger than $\theta_{tc0}$ will be confined in the slab; thus no refracted light could travel out. At this incident angle, there is no reflection light because it is larger than the critical angle of reflection $\theta_{rc0}$. One can easily obtain that $\theta_{rc0} = 90° - 2\theta_0$ by setting the reflected light parallel to the boundary $y' = y = -h$ ($x \geq 0$) so as to confine the light in the slab. Above we have set $0° \leq \theta_0 \leq 90°$ without loss of generality, if $-90° \leq \theta_0 \leq 0°$ (i.e., another kind of coordinate compression), then $\theta_{tc0} = -90° - \theta_0$ and $\theta_{rc0} = -90° - 2\theta_0$. What is more, apart from the above two critical angles, if $n_1 > n_2$, there should also be another two:

$$\theta_{tc} = \arcsin(\pm \frac{n_2}{n_1}) - \theta_0, \tag{5}$$

which share the same physics of the conventional critical angles. By looking into the modified laws, it is not difficult to find cases of negative refraction and negative reflection, which also exist in the cases of phase discontinuities. For example, if $\theta_i < 0$ (i.e., incident from the fourth quadrant) and ($|\theta_i| < \theta_0$), there should be positive $\theta_t$, as a negative refraction. Likewise, if $\theta_i < 0$ and ($|\theta_i| < 2\theta_0$), there should be positive $\theta_r$, as a negative reflection.

In the following, we will investigate the relationship of the incident angle with the refraction angle and that with the reflection angle. As known, there are two ways that light passes though the interface between different media. One is light travelling from an optically thinner medium to an optically denser medium. The other is light travelling from an optically denser medium to an optically thinner medium. What would happen under these modified laws of refraction and reflection in the two different ways? The above four (or two, if choose

$n_1 < n_2$) critical angles will divide the range of incident angle $[-90°, 90°]$ into different regions, where different interesting phenomena occur. We set $k = \tan\theta_0 = 0.25$ ($\theta_0 = 14°$). The angles of refraction and reflection as functions of the incident angle $\theta_i$ are shown in Fig. 2(a) (for $n_1 = 1$ and $n_2 = 2$) and 2(b) (for $n_1 = 2$ and $n_2 = 1$), in which curves at different intervals reflect different phenomena on the interface. In both figures, curves that located in the second quadrant denote negative refraction and negative reflection. In Fig. 2(a), light propagates from optically thinner medium to optically denser medium. For $-14° \leq \theta_i \leq 0°$, the meta-surface exhibits negative refraction. For $62° \leq \theta_i \leq 76°$, the reflected light is totally confined in the meta-surface while the refracted light still propagates. For $76° \leq \theta_i \leq 90°$, the reflected light and the refracted light are both confined in the meta-surface. The light will propagate along the meta-surface just like evanescent waves. The angle $\theta_{rc0}$ and $\theta_{tc0}$ will not change accordingly when the configuration is modified that light incident from an optically denser medium to an optically thinner medium as they are intrinsically born from the transformation itself, see in Fig. 2(b). However, according to Eq. (5), there will be two more critical angles come out, that is $\theta_{tc} = 16°$ and $-43°$. For $-90° \leq \theta_i \leq -43°$ and $16° \leq \theta_i \leq 62°$, no refraction occurs. While for $62° \leq \theta_i \leq 90°$, neither refraction nor reflection occurs. All light will be localized in the mete-surface. Therefore, the two critical angles from the transformation geometry enrich the physics of the laws of reflection and refraction, rendering different sequences of the above four critical angles and related phenomena.

To show the phenomenon more visually, we perform numerical simulations using the finite element solver COMSOL Multiphysics. We set $n_1 = 1$, $n_2 = 2$, $k = \tan\theta_0 = 0.15$ ($\theta_0 = 8.53°$), $h = 0.02$ (a.u.). As the electromagnetic wave can always be decoupled into a transverse electric (TE) polarized wave (electric field along $z$ direction) and a transverse magnetic (TM) polarized wave (magnetic field along $z$ direction), we only consider the TM modes during the simulations. The incidence is chosen as a Gaussian beam, and the wavelength is set to be 0.3, which is 15 times of the width of the slab so that it can be regarded as a meta-surface. Figure 3(a) is the magnitude of magnetic field distribution for ordinary reflection and refraction. Light is incident from a dielectric with $n_1 = 1$ to another with $n_2 = 2$, without adding any slab at the interface. As seen, the incident and reflection angle are both 20° while the refraction angle is 9.85° (obeying the conventional laws). Figure 3(b) shows the magnitude of magnetic field distribution for anomalous reflection and refraction generated by the transformation media slab added at the interface. The incident angle is 20° as well for comparison, while the reflection angle is now changed into $\theta_r = 37°$ and the refraction angle into $\theta_t = 13.8°$ (obeying the modified laws).

In conclusion, we have shown that transformation optics can also be utilized to design meta-surface that can change the laws of reflection and refraction. The material parameters of transformation optical "surface" are complicated thus difficult to implement. However, seeing the recent progress of the invisibility cloaks [5-10], we still look forward to a promising future for this technique. The physics of manipulating phases behind the concept of phase discontinuities and transformation optics are different. The former one uses local resonances of pre-designed antennas, hence the operating slab is ultra-thin and more sub-wavelength. The latter uses modulated refractive indexes to achieve parallel phenomena, thus providing additional insight and exploration to the concept.

This work was supported by the National Natural Science Foundation of China (grant no. 11004147), the Natural Science Foundation of Jiangsu Province (grant no. BK2010211) and the Priority Academic Program Development (PAPD) of Jiangsu Higher Education Institutions.

**Figures:**

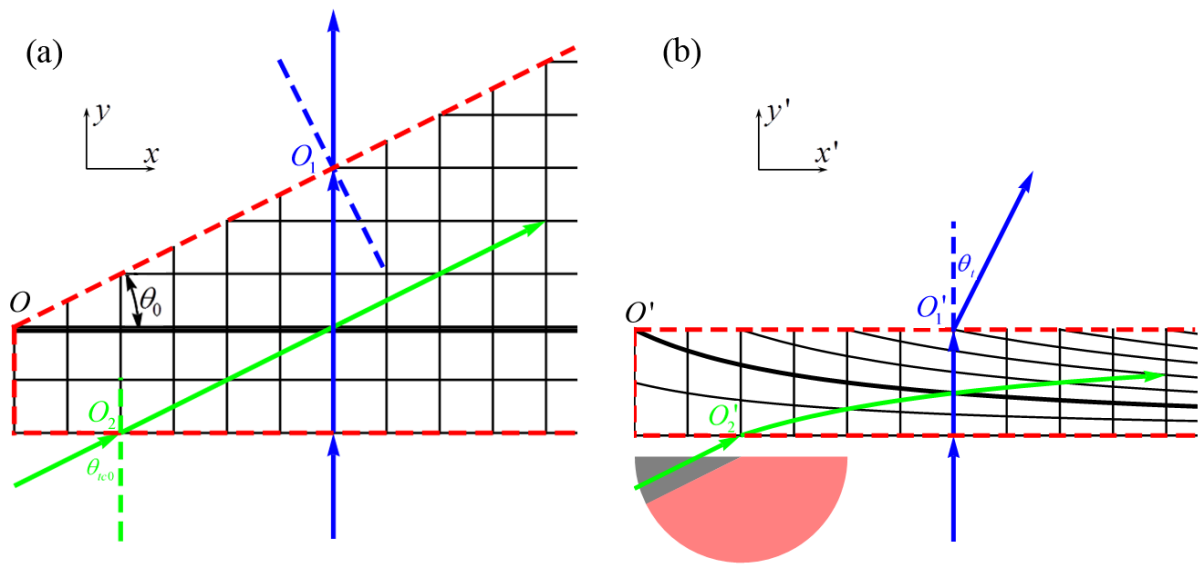

Fig. 1(Color online). (a) The virtual space, a trapezoidal region bounded by red/dashed lines. The lower boundary is $y=-h$ $(x\geq 0)$ and the upper one is $y=kx=x\tan\theta_0$ $(x\geq 0)$. The light propagates in straight rays (the green/gray and blue/black arrow lines) in a homogeneous dielectric $n_1$. (b) The physical space, a slab with a width $h$ bounded by red/dashed lines. The light propagates in bended or squeezed rays (the green/gray and blue/black arrow curves) in the transformation media slab. The semicircle signifies that for some angles of incidence (dark gray part) the light cannot reach the upper boundary of the slab, indicating a critical angle of refraction directly from the transformation.

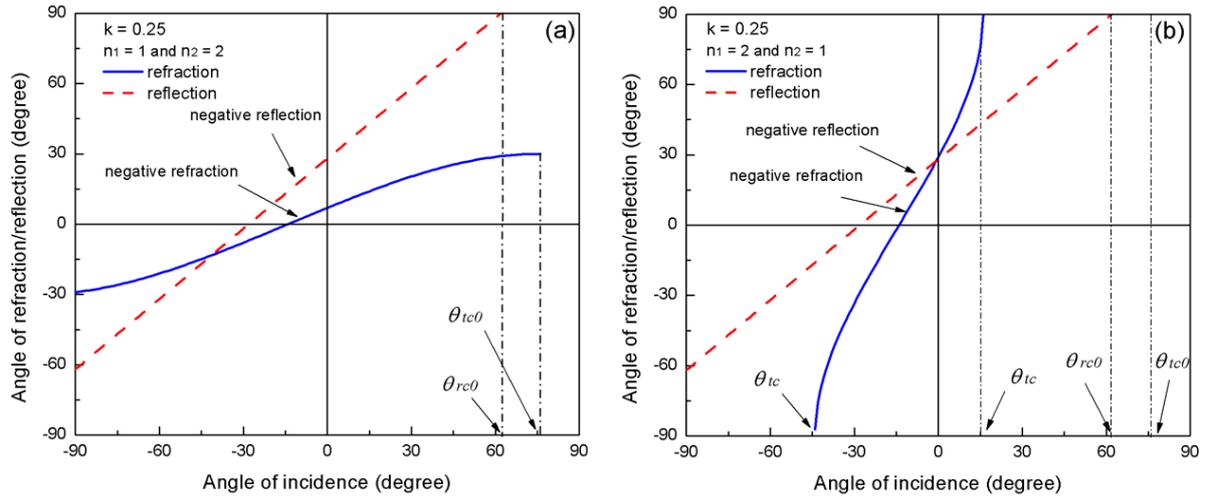

Fig. 2(Color online). The relationship between the angle of refraction/reflection and the incident angle. The curves in the second quadrant denote negative refraction and reflection. (a) is the case that light incident from an optically thinner medium to an optically denser medium. (b) is the case that light incident from an optically denser medium to an optically thinner medium.

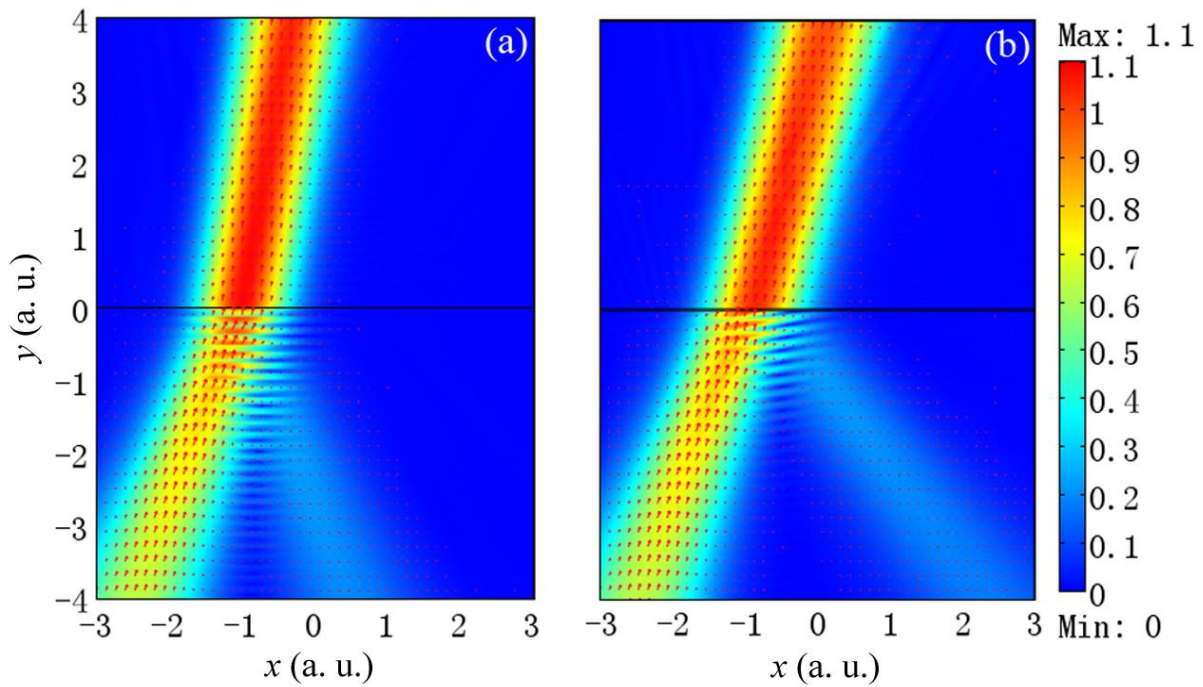

Fig. 3(Color online). (a) The magnitude of magnetic field for ordinary reflection and refraction. (b) The magnitude of magnetic field for anomalous reflection and refraction.